\documentclass[sn-mathphys-num]{sn-jnl}

\usepackage{graphicx}%
\usepackage{multirow}%
\usepackage{amsmath,amssymb,amsfonts}%
\usepackage{amsthm}%
\usepackage{mathrsfs}%
\usepackage[title]{appendix}%
\usepackage{xcolor}%
\usepackage{textcomp}%
\usepackage{manyfoot}%
\usepackage{booktabs}%
\usepackage{algorithm}%
\usepackage{algorithmicx}%
\usepackage{algpseudocode}%
\usepackage{listings}%
\usepackage{float}
\usepackage{algorithm}
\usepackage{algpseudocode}

\theoremstyle{thmstyleone}%
\theoremstyle{thmstyletwo}%

\theoremstyle{thmstylethree}%

\begin{document}

\title[Article Title]{A Numerical Investigation of Particle Deposition on a Substrate}


\author[1]{\fnm{Ananta Kumar}\sur{Nayak}}

\author[1]{\fnm{Amritpal}\sur{Singh}}

\author[2]{\fnm{Mehrdad}\sur{ Mesgarpour}}
\author*[1,3]{\fnm{Mostafa Safdari}\sur{Shadloo}}\email{msshadloo@coria.fr}
\affil[1]{INSA Rouen Normandie, Univ Rouen Normandie, CNRS, Normandie University, CORIA UMR 6614, F-76000 Rouen, France}

\affil[2]{School of Business, Society and Engineering, Mälardalens University, 72220, Västerås, Sweden}

\affil[3]{Institut Universitaire de France, Rue Descartes, F-75231 Paris, France}


\abstract{The deposition of nanometer-scale particles is of significant interest in various industrial processes. While these particles offer several advantages, their deposition can have detrimental effects, such as reducing the heat transfer efficiency in nanofluid-based battery cooling systems. In this study, we investigated particle deposition around different square substrate configurations as well as experimentally obtained complex porous structure in a two-dimensional setup. The particles modeled as a concentration field using the lattice Boltzmann method, with a given external flow following a parabolic profile. Our results revealed that particle deposition around a substrate increases with higher fluid velocity, greater particle concentration, and higher deposition probability. Additionally, placing multiple number of substrates in the channel resulted in increased deposition on upstream substrates compared to downstream ones. As particle deposition around upstream substrates increases, it eventually obstructs the flow to downstream regions, thereby affecting the overall system performance. The insights gained from this simplified model of particle deposition will play a crucial role in advancing our understanding of deposition processes in complex systems.}




\maketitle

\section{Introduction}
\label{intro}
Nanofluids, which contain nanometer-sized particles suspended in a carrier fluid, are widely used in industrial applications such as battery cooling of electric vehicles due to their higher thermal conductivity and surface area-to-volume ratio \cite{chakraborty2020stability}. However, these particles over time tend to form clusters, which can lead to sedimentation on the heat transfer surfaces and reduce heat transfer efficiency due to increased thermal resistance. In addition to cooling systems, other industrial devices such as boilers, furnaces, mufflers, and thermoelectric equipment require treatment of the flue gases released during their use \cite{tong2017simulation}. These flue gases contain harmful particulate matters (PMs), CO$_2$, NO$_x$, and SO$_2$, affecting the environment as well as human health \cite{omidvarborna2015recent, gharehghani2021applications}. With stricter emissions regulations proposed by various countries to mitigate global warming, health risks, and natural disasters, the treatment of flue gases before their release into the atmosphere is now inevitable \cite{omidvarborna2015recent, jiaqiang2022soot}.

The capture of PM is of particular concern because these particles can accumulate over time, causing blockages in the filter designed to capture them \cite{gharehghani2021applications, jiaqiang2022soot}. These blockages not only reduce the efficiency of heat transfer by impeding the flow of gases but also increase pressure drops within the system, resulting operational issues, including higher energy consumption. 
The filters used for PM capture are generally porous honeycomb structures, which can be constructed by placing posts or obstacles aligned in the direction of the gas flow \cite{yamamoto2013numerical}. The efficacy of a filter depends on several factors, such as the size and shape of the posts or substrates and their arrangement. For example, smaller pores that allow flue gases to pass through result in better filtration but lead to higher pressure drops \cite{zuccaro2011multiphase}. Therefore, a trade-off between filtration efficiency and pressure drop is necessary.
Furthermore, the dynamics of particles play a crucial role in determining how they tend to form clusters and subsequently deposit on the substrate \cite{wang2024numerical}.

  Although recent studies have focused primarily on enhancing heat transfer through the use of nanofluids, a comprehensive analysis of their long-term effects on heat transfer remains unclear. It is well known that over time, nanoparticles can accumulate and form sediments on inner surfaces, which can affect the heat transfer efficiency of a system \cite{evans2008effect, chakraborty2020stability}. Consequently, further research is required to understand the physics of particle deposition, from the initial stages to the development of larger clusters that can ultimately obstruct fluid flow and hinder heat transfer.  
This aspect remains largely unexplored in the literature because of the challenges associated with conducting experiments and tracking particle deposition over longer timescales than those involved in cluster formation. To overcome these challenges, numerical methods have been widely adopted over the last three decades \cite{yamamoto2013numerical, kong2019simulation}. Most studies have focused on the impact of fluid dynamics on nanoparticle deposition. For example, Ansari et al. \cite{ansari2015numerical} investigated the roles of various forces, including Brownian motion, lift, drag, and gravity, in the deposition of nanoparticles on substrates in gaseous fluids. Additionally, investigations into particle deposition have mainly aimed at identifying the forces responsible for their deposition and understanding their effects on the heat transfer enhancement of fluids \cite{albojamal2017analysis, mazaheri2021two}. 
Despite the use of numerical simulations, these studies predominantly focus on short-term effects, such as heat transfer analysis, due to the computational costs associated with simulating millions of particles over extended timescales.

  Due to the limited access to high computational power for simulating millions of particles, simplified deposition models have been proposed. For instance, Yamamoto et al. \cite{yamamoto2009simulation} represent particles not as solids flowing within a pipe but rather as a concentration field. This approach enables the qualitative analysis of particle deposition phenomena over large timescales while requiring limited computational power. Furthermore, the deposition of particles using this approach is numerically simulated using the lattice Boltzmann method (LBM). Over the past few decades, the LBM has emerged as a powerful numerical solver and an alternative to conventional methods, such as the finite volume method (FVM), finite difference method (FDM), and finite element method (FEM), particularly for pore-scale transport processes \cite{wang2013lattice, zhou2015lattice}. The LBM offers several important advantages, including ease of implementation, high parallelism, and the ability to deal with complex physical phenomena and boundary conditions \cite{alamian2024modeling}. Therefore, LB models have been developed in the past to simulate physio-chemical phenomena, such as dissolution, precipitation, adsorption, and combustion processes \cite{zhou2015lattice,  stockinger2024lattice, chen2014pore}. It is important to note that simulating complex reactive transport processes with evolving solid structures typically involves coupling transport processes with models that track dynamically evolving fluid-solid interfaces \cite{stockinger2024lattice}. In this regard, the LBM is particularly well-suited for capturing these complexities, providing enhanced capabilities for simulating pore-scale phenomena.

  Most numerical studies \cite{li2024study, huang2024study, zhou2016lattice} have primarily focused on particle deposition or gas adsorption on substrates with complex topologies. Consequently, the underlying physics, such as the influence of flow fields, substrate geometry and its arrangement on local particle deposition, are still not fully understood. While a few studies \cite{zhou2015lattice, ansari2015numerical, tong2017simulation} have examined particle deposition on simple substrates, such as cylindrical or square geometries, to clarify the mechanisms of local deposition, comprehensive studies remain lacking. These would help to fully explain the factors influencing particle deposition and the impact of evolving fluid-solid layers on flow velocity across different substrate arrangements within a channel.

  In this study, we employ the LBM to investigate the deposition of solute concentration and the subsequent formation of layers around a simple square substrate in two-dimensional flow. We further explore how the arrangement of multiple square substrates influences the flow field and how this, in turn, affects particle deposition around the solid substrates. Additionally, we investigate particle deposition in a complex porous network obtained from an experiment using Focused Ion Beam Scanning Electron Microscopy (FIB-SEM). Furthermore, we investigate the effects of varying flow velocity, inlet solute concentration, and deposition probability on the evolving structure around the substrate.

\section{Methodology}
\label{methods}
\subsection{Governing equations}
\begin{figure}
  \centering
  \includegraphics[width=\textwidth]{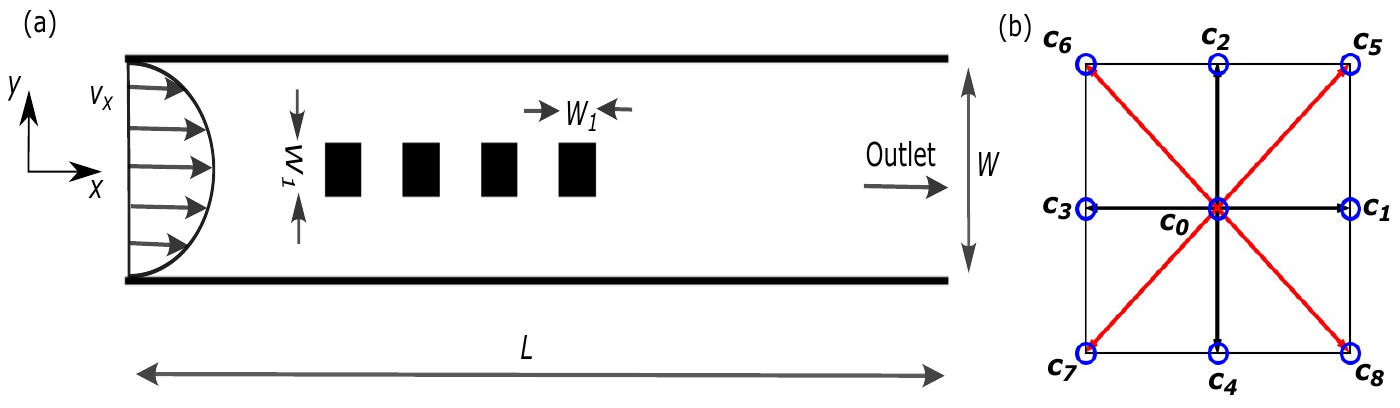}
  \caption{Schematic shows: (a) a computational setup consisting a straight channel of length $L$ and width $W$, with four square substrates each having an area of $W_1 \times W_1$, arranged in tandem along the center of the channel. A parabolic flow is applied at the inlet, positioned at the left end of the channel; (b) the direction discrete velocity vectors of the D2Q9 model.}
  \label{computational_domain}
  \end{figure}

In this study, we investigate the flow of the particle mass fraction concentration and its subsequent deposition on square substrates with dimensions $W_1 \times W_1$  within a two-dimensional channel of length $L$ and width $W$. Our investigation begins with a single substrate placed at the center of the channel and further extends to configurations with multiple substrates. For example, we will also analyze the deposition of particles on four square substrates arranged in tandem at the center of the channel, as shown in Figure \ref{computational_domain}(a). This particle mass fraction concentration field evolves with time in the channel and is obtained by solving the advection-diffusion-reaction equation, which is represented mathematically as follows
\begin{equation}
		\frac{\partial Y_c}{\partial t} + \nabla \cdot (Y_c \mathbf{v}) = \nabla \cdot (D_c   \nabla Y_c) \pm F_s \,,
  \label{mass_fraction_soot}
\end{equation}

  where $F_s$ is the source/sink term, and the sign $\pm$ before $F_s$ in equation \eqref{mass_fraction_soot} indicates a source ($+$) or sink ($-$), which is defined as
\begin{equation}
    F_s = \frac{D_c}{\rho} \nabla Y_c \cdot \nabla \rho + Y_c \nabla \cdot\mathbf{v} \,,
    \label{source_sink_term}
\end{equation}
where $\mathbf{v}$ and $\rho$ denote fluid velocity and density fields, and $Y_c$ and $D_c$ represent particle mass fraction and diffusion coefficient, respectively.

  The velocity and density fields in the computational domain are obtained by solving the Navier-Stokes equations, 
	\begin{equation}
		\frac{\partial \rho}{\partial t} + \nabla \cdot (\rho \mathbf{v}) = 0\,,
		 \label{continuity equation}
	\end{equation}
	\begin{equation}
		\frac{\partial (\rho \mathbf{v})}{\partial t}+ \nabla \cdot (\rho \mathbf{v} \mathbf{v}) = -\nabla p + \nabla.(\mu \nabla \mathbf{v}) + \mathbf{F}\,,
		 \label{momentum_equation}
	\end{equation}
where $\mu$ is the dynamic viscosity of the fluid, $p$ is the pressure field, and $\mathbf{F}$ is an arbitrary volumetric force vector field acting onto the flow. We impose a parabolic velocity profile at inlet of the channel, which is given as 

	\begin{equation}
        v_x = v_{\text{max}}\left[1-\left(\frac{y}{W/2}\right)^2 \right],
		 \label{VX_equation}
	\end{equation}
where $v_{\text{max}}$ is the maximum flow velocity imposed at the channel center, and $W$ is the channel width. The velocity component perpendicular to the flow direction is $v_y = 0$. We initially solve for steady flow in the computational domain by solving equation \eqref{momentum_equation} with the inlet flow condition imposed by equation \eqref{VX_equation}, no-slip boundary conditions at solid walls, and a zero x-component velocity-gradient condition at the channel outlet. After obtaining the steady flow velocity profile in the channel, we solve for the particle mass fraction concentration, i.e. equation \eqref{mass_fraction_soot}, for a given inlet concentration. In addition, we impose a no-flux boundary condition on solid walls including substrates.
\subsection{Numerical method}
We employ the widely used single relaxation time-LBM to solve equations \eqref{mass_fraction_soot} and \eqref{momentum_equation}. In LBM, two key steps, collision and streaming, are used. Unlike other numerical methods that solve directly for macroscopic quantities, the LBM derives velocity and density from particle dynamics via a probability distribution function. These particles move along a predefined Cartesian grid, known as lattice nodes, with the D2Q9 model commonly used. This model represents two-dimensional space with nine possible discrete velocity directions for particle movement from each lattice node, as shown in Figure \ref{computational_domain}(b).
Mathematically, the time evolution of distribution function $f_i(\mathbf{x},t)$ for fluid is given as \cite{{mohamad2011lattice}}
\begin{equation}
 f_{i}(\mathbf{x} + \mathbf{c}_{i} \Delta t,t+\Delta t) - f_i(\mathbf{x},t) = \Delta t (\Omega_i + F_i) \,, 
\label{probility_distributuion_function}
\end{equation}
where $\mathbf{c}_i$ is the discrete lattice velocity vector for the distribution function $i$, with $i$ ranging from 0 to 8. The Bhatnagar-Gross-Krook (BGK) collision operator $\Omega_i$ is represented as follows,
\begin{equation}
\Omega_i = - \frac{1}{\tau}\left[f_i(\mathbf{x},t) - f^{eq}_i(\mathbf{x},t)\right] \,,
\label{BGK_collision_operator}
\end{equation}
where $\tau$ is the relaxation time, which is related to the kinematic viscosity of the fluid. The value of $\tau$ is considered 1 in all simulations \cite{stockinger2024lattice}. The physical kinematic viscosity of the fluid, while obtaining Navier-Stokes equations from the lattice Boltzmann equation by employing the Chapman-Enskog multi-scale approach, is derived as \cite{mohamad2011lattice} 
\begin{equation}
\nu = \frac{1}{2} c^2_s \Delta t (2 \tau -1) \,,
\label{kinematic_viscosity}
\end{equation}
where $c_s = \frac{1}{\sqrt{3}} \frac{\Delta x}{\Delta t}$ is the sound speed. The right hand side of equation \eqref{probility_distributuion_function} represents the collision step, whereas left hand side refers to the streaming step. The discretized form of the equilibrium distribution function, $f_i^{\text{eq}}$, is obtained from as an approximation of the Maxwell probability function. This is expressed as the functions of local lattice and macroscopic velocity of the fluid,
\begin{equation}
f_i^{\text{eq}} = \omega_i \rho \left[1+\frac{1}{c_s^2}(\mathbf{c}_i \cdot \mathbf{v})+ \frac{1}{2c_s^4}(\mathbf{c}_i \cdot \mathbf{v})^2-\frac{1}{2c_s^2}(\mathbf{v} \cdot \mathbf{v})\right] \,.
\label{equilibrium_distribution_function}
\end{equation} 
The nine discrete velocity vectors used in the D2Q9 model is given as, 
\begin{equation}
\mathbf{c}_i = \begin{cases}
(0,0), & i = 0 , \\
(\text{cos}[i-1]\frac{\pi}{2}, \text{sin}[i-1]\frac{\pi}{2})\frac{\Delta x}{\Delta t},  & i = 1, 2, 3,4 , \\
(\text{cos}[2i-9]\frac{\pi}{2}, \text{sin}[2i-9]\frac{\pi}{2})\frac{\Delta x}{\Delta t},  & i = 5, 6, 7, 8,
\end{cases}
\label{velocity_vectors}
\end{equation}
where $\Delta x$ and $\Delta t$ are the spatial and temporal discretization steps, respectively. The weight $\omega_i$ used in the D2Q9 model is given by,
\begin{equation}
\omega_i = \begin{cases}
\frac{4}{9}, & i = 0 , \\
\frac{1}{9},  & i = 1, 2, 3,4 , \\
\frac{1}{36},  & i = 5, 6, 7, 8.
\end{cases}
\label{weight_factors}
\end{equation} 
To account for volumetric forces affecting fluid flow, such as gravity, the discrete form of the bulk force, $F_i$ proposed by Guo and Zhao \cite{guo2002lattice},  as follows 
\begin{equation}
F_i = \left(1-\frac{1}{2\tau}\right) \omega_i \left(\frac{\mathbf{c}_i - \mathbf{v}}{c^2_s} + \frac{\mathbf{c}_i \cdot \mathbf{v}}{c^4_s} \mathbf{c}_i\right) \mathbf{F} \,.
\end{equation}
The macroscopic density and velocity at each lattice nodes are obtained zeroth and first moments of the distribution function as follows,
\begin{equation}
\rho(\mathbf{x},t + \Delta t) = \sum_{i = 0}^{8}  f_{i}(\mathbf{x},t + \Delta t)\,,
\label{macroscopic_fluid_variable_density}
\end{equation}
\begin{equation}
\rho(\mathbf{x},t + \Delta t) \mathbf{v}(\mathbf{x},t + \Delta t) = \sum_{i = 0}^{8} \mathbf{c}_i f_{i}(\mathbf{x},t + \Delta t) + \frac{1}{2} \Delta t F_i(\mathbf{x},t)\,.
\label{macroscopic_fluid_variable_velocity}
\end{equation}
For the solution of equation \eqref{mass_fraction_soot}, we solve for the time evolution of mass fraction of the particle distribution function $g_{c,i}(\mathbf{x},t)$, which is given as follows,
\begin{equation}
 g_{c,i}(\mathbf{x} + \mathbf{c}_{i} \Delta t,t+\Delta t) - g_{c,i}(\mathbf{x},t) = \Delta t \left(\Omega_{c,i} +  F_{s,i}  + \frac{1}{2} \Delta t \frac{\partial F_{s,i}}{\partial t} \right) \,, 
\label{soot_probility_distributuion_function}
\end{equation}
where  BGK collision operator $\Omega_{c,i}$ is expressed as follows
\begin{equation}
\Omega_{c,i} = - \frac{1}{\tau_c}\left[g_{c,i}(\mathbf{x},t) - g^{eq}_{c,i}(\mathbf{x},t)\right] \,,
\label{soot_BGK_collision_operator}
\end{equation}
where $\tau_c$ is the single relaxation time (SRT), derived as a closure condition while obtaining equation \eqref{mass_fraction_soot} from the lattice Boltzmann equation through Chapman-Enskog analysis, and is given as,
\begin{equation}
D_c = \frac{1}{2} c^2_s \Delta t (2 \tau_c -1) \,.
\label{diffusion_coefficient_viscosity}
\end{equation}
 The particle mass fraction diffusion coefficient $D_c$ is taken 0.1562 in lattice unit
 \cite{stockinger2024lattice}. The discrete form of the equilibrium distribution function $g_{c,i}^{\text{eq}}$ is defined as
\begin{equation}
g_{c,i}^{\text{eq}} = \omega_i Y_c(\mathbf{x},t) \left[1+\frac{1}{c_s^2}(\mathbf{c}_i \cdot \mathbf{v})+ \frac{1}{2c_s^4}(\mathbf{c}_i \cdot \mathbf{v})^2-\frac{1}{2c_s^2}(\mathbf{v} \cdot \mathbf{v})\right] \,.
\label{soot_equilibrium_distribution_function}
\end{equation} 
The discrete form of the source or sink term is expressed as,
\begin{equation}
F_{s,i} = \omega_i F_s \left[1  + \frac{\mathbf{c}_i \cdot \mathbf{v}}{c^2_s } \left(1 -\frac{1}{2\tau_c}\right) \right] \,.
\end{equation}
The mass fraction of particle at lattice nodes is obtained as follows,
\begin{equation}
Y_c(\mathbf{x},t + \Delta t) = \sum_{i = 0}^{8} g_{c,i}(\mathbf{x},t + \Delta t)\,.
\end{equation}
The boundary conditions for fluid and mass fraction of particle are implemented using bounceback method as elaborated in \cite{stockinger2024lattice}.

\subsection{Particle deposition model}
There are several Lagrangian models used to simulate the particle deposition. However, simulating millions of nanometer-sized particles and tracking them over long periods using the Lagrangian approach is a cumbersome task. Instead, we solve for the mass concentration of particles in the computational domain. The mass fraction of particles deposited on the substrate surface is determined using the following equation \cite{yamamoto2009simulation}
\begin{equation}
    Y_{c,s}(\mathbf{x},t+\Delta t) =  Y_{c,s}(\mathbf{x},t) + DP \sum_{i = 0}^{8} g_{c,i}(\mathbf{x},t + \Delta t) \,,
    \label{surface_deposition}
\end{equation}
where $Y_{c,s}$ and $DP$ represent the mass fraction of particles and the probability of particle deposition on the substrate surfaces, respectively. As particles accumulate in the gas phase at the substrate surface over time, $Y_{c,s}$ increases. Once particle deposition reaches unity, the nearest fluid nodes to the surface become solid nodes. These deposited solid nodes are then treated as no-slip boundary conditions, leading to a dynamic change in both the fluid and mass fraction of particle boundary conditions.
\subsection{Interface node identification algorithm for particle deposition}
First, we assign numerical tags to the computational domain: 0 for fluid nodes, 1 for solid walls, 2 for substrate nodes, and 3 for deposited solid nodes. Next, we utilize these tags along with the D2Q9 discrete lattice velocity vectors to identify fluid nodes adjacent to the substrates. If any of the nine streaming positions in the D2Q9 directions correspond to a substrate node, we calculate the mass fraction of deposited particles for that fluid (or interface) node over time using equation \eqref{surface_deposition}. When the mass fraction of deposited particles exceeds unity, the fluid node is re-tagged as 3.
\section{Results and discussion}
\label{results_and_discussion}
In this study, we apply the LBM to investigate particle deposition on a substrate. This method provides a simplified framework for exploring how particle deposition affects the fluid field around the substrate and , in turn, how changes in fluid flow influence particle deposition. The LBM algorithm used in this study is detailed in the following pseudo-algorithm steps as shown in Algorithm \ref{alg:particle_deposition}. To focus on the underlying physics following deposition, we simulate particle accumulation on various simple geometric configurations. The insights gained from this approach can guide the development of more advanced models and controlled techniques for particle deposition in complex geometries. Since the problem involves a complex interplay between fluid flow, particle mass fraction dynamics, and dynamic boundary changes around substrates, therefore, we simplify the problem by considering several key assumptions in our study.
\begin{algorithm}
\caption{Simulation LBM Algorithm with Boundary Conditions and Deposition}\label{alg:particle_deposition}
\begin{algorithmic}
\State Initialize macro variables
\While{stop criterion not reached}
    \State Check stop criterion
        \If{stop criterion is met}
            \State Terminate process
        \EndIf
        \State Perform the collision step
        \State Perform the streaming step
        \If{node is a solid node}
            \State Apply the bounce-back boundary condition
        \EndIf
        \If{Substrate surface node deposition criterion is met}
            \State Update the geometry
        \EndIf
\EndWhile
\end{algorithmic}
\end{algorithm}


  Firstly, we do not specifically track the motion of individual Lagrangian particles prior to deposition. Instead, we represent the solid particles as a concentration field, which reduces the computational demand for simulating millions of nanoparticles. Secondly, we assume the substrate has an inherent ability to attract the particles, allowing them accumulate and eventually solidify. Thirdly, we assume that once solid particles are deposited, they have the same potential to attract other particles, leading to self-sustained deposition growth over time, when the mass fraction of the concentration field at the interface reaches unity. Furthermore, thermodynamic effects related to phase changes during deposition are not taken into account in this study.

  In the following subsection, we simulate the deposition of the particle concentration field onto solids at lattice nodes. We simulate particle deposition by changing the flow strength, deposition probability, and inlet mass fraction of the particle concentration within a straight channel having different configurations of square substrates. To quantify the effects of variations in flow strength, deposition probability, and inlet mass fraction on particle deposition and changes in velocity within the computational domain, we define the following dimensionless numbers. 
$$\text{Péclet number}  (Pe) = \frac{{v_\text{avg,x}} W_1}{D_c}, \quad \text{Dimensionless time} (T^\star) = \frac{t}{t_0},$$  $$\quad \text{Deposition thickness to substrate size ratio (DT-SSR)} = \frac{\text{Deposition thickness}}{W_1},$$
where $v_\text{avg,x}$ is the average inlet velocity, and $t_0$ represents the total simulation time (is 1.35e+06). Furthermore, we used the code developed by Stockinger et al. \cite{stockinger2024lattice}, which has been validated with various test cases, in our investigation. The size of the computational domain is 420 $\times$ 80 lattice units, with each square substrate having dimensions of 20 $\times$ 20 lattice units. In this two-dimensional study, the lattice spacing is considered as 1.002 $\mu$m.
\subsection{Particle deposition on single substrate}
In the first case, we evaluate particle deposition on a substrate, which is placed 230 lattice units from the inlet of the channel. Fluids containing a concentration of particles are introduced at different velocities. In addition, particle deposition is also influenced by both the $DP$ and the mass concentration of the particles ($Y_c$) in the gas. To understand the influence of these variables on particle deposition, we conduct different simulations to assess the effects of $DP$, $Y_c$, and $Pe$.
\subsubsection{Effect of flow velocity and deposition probability}
\begin{figure}[!hbt]
\centering 
    \begin{minipage}[b]{0.8\textwidth}
        \centering
        \includegraphics[clip, trim=0 0 0 0, width=\textwidth]{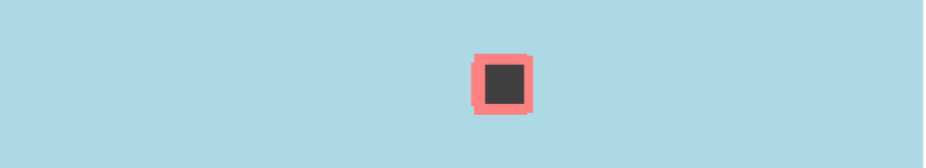} 
        (a)
    \end{minipage}
   \vspace{0.5cm} 
    \begin{minipage}[b]{0.8\textwidth}
        \centering
        \includegraphics[width=\textwidth]{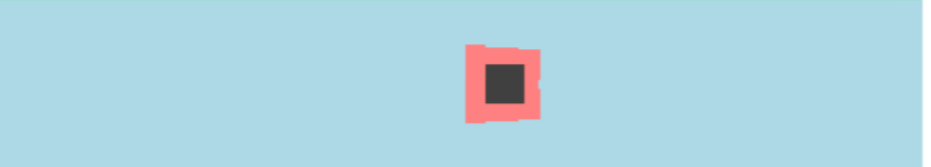} 
        (b)
    \end{minipage}
     \caption{Simulation snapshots show particle deposition around the substrate at time $T^\star$ = 0.2  for $Pe = 0.01$ and inlet mass fraction of concentration $Y_c =0.01$, comparing two cases of particle deposition probability ($DP$): (a) 0.006 and (b) 0.01.}
    \label{fig:result1}
\end{figure}

Figure \ref{fig:result1} shows the contour of particle deposition on the substrate at $Pe = 0.01$ for two different $DP$, 0.006 and 0.01, visualized at time $T^\star=0.2$. Here, $T^\star$ represents the ratio of simulation time to total time. It is evident that as $DP$ increases from 0.006 (Figure \ref{fig:result1}(a)) to 0.01 (Figure \ref{fig:result1}(b)), particle deposition on the substrate increases. Additionally, deposition around the substrate surface at $DP=0.01$ is more asymmetrical than at $DP=0.006$. The increased deposition at $DP=0.01$ results from higher conversion of concentration $Y_c$ on the substrate, leading to faster deposition, as associated in equation \eqref{surface_deposition}.

\begin{figure}[!hbt]
    \begin{minipage}[b]{0.5\textwidth}
        \centering
        \includegraphics[clip, trim=0 0 0 0, width=\textwidth]{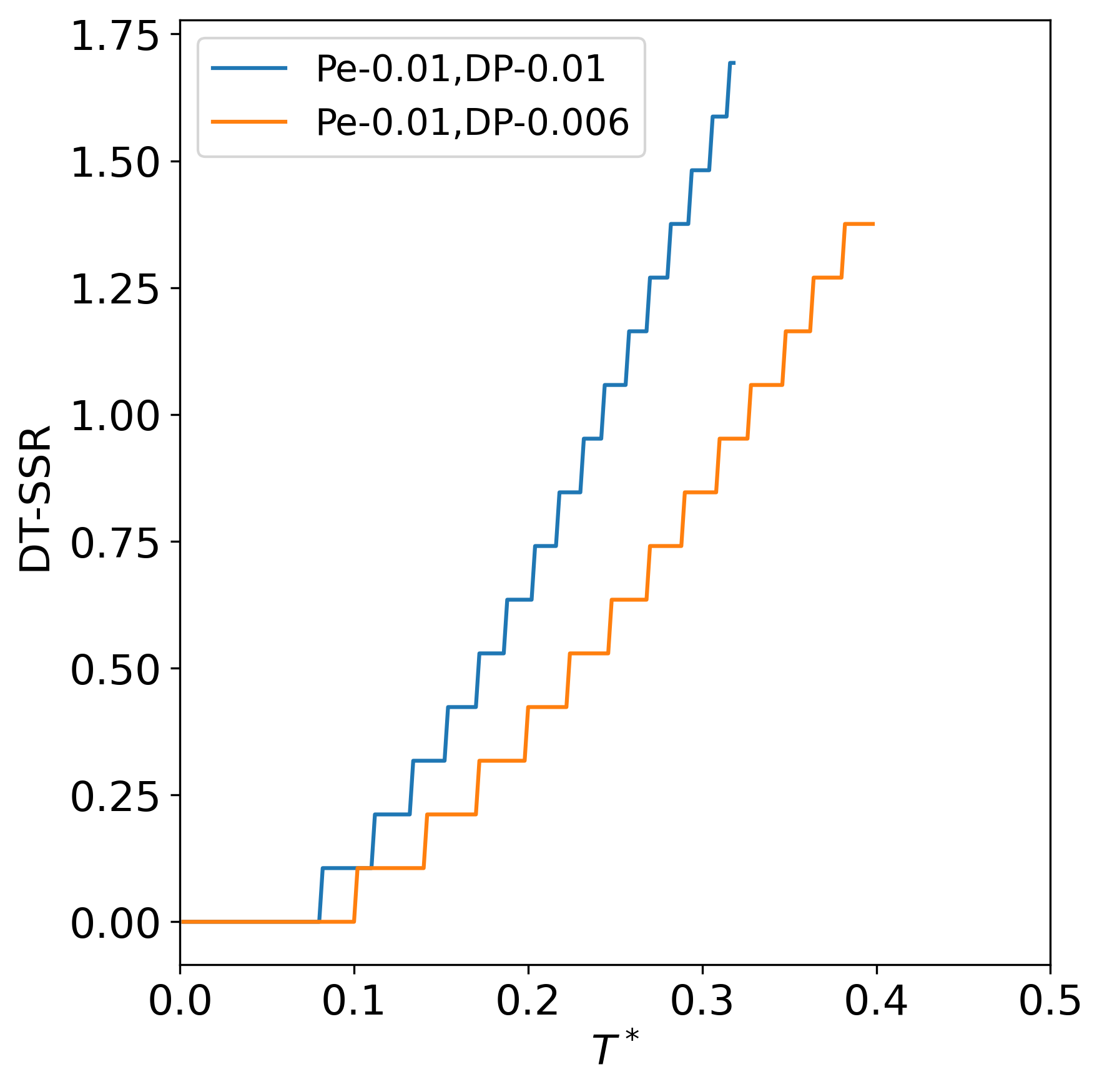} 
        (a)
    \end{minipage}
    \hfill 
    \begin{minipage}[b]{0.5\textwidth}
        \centering
        \includegraphics[width=\textwidth]{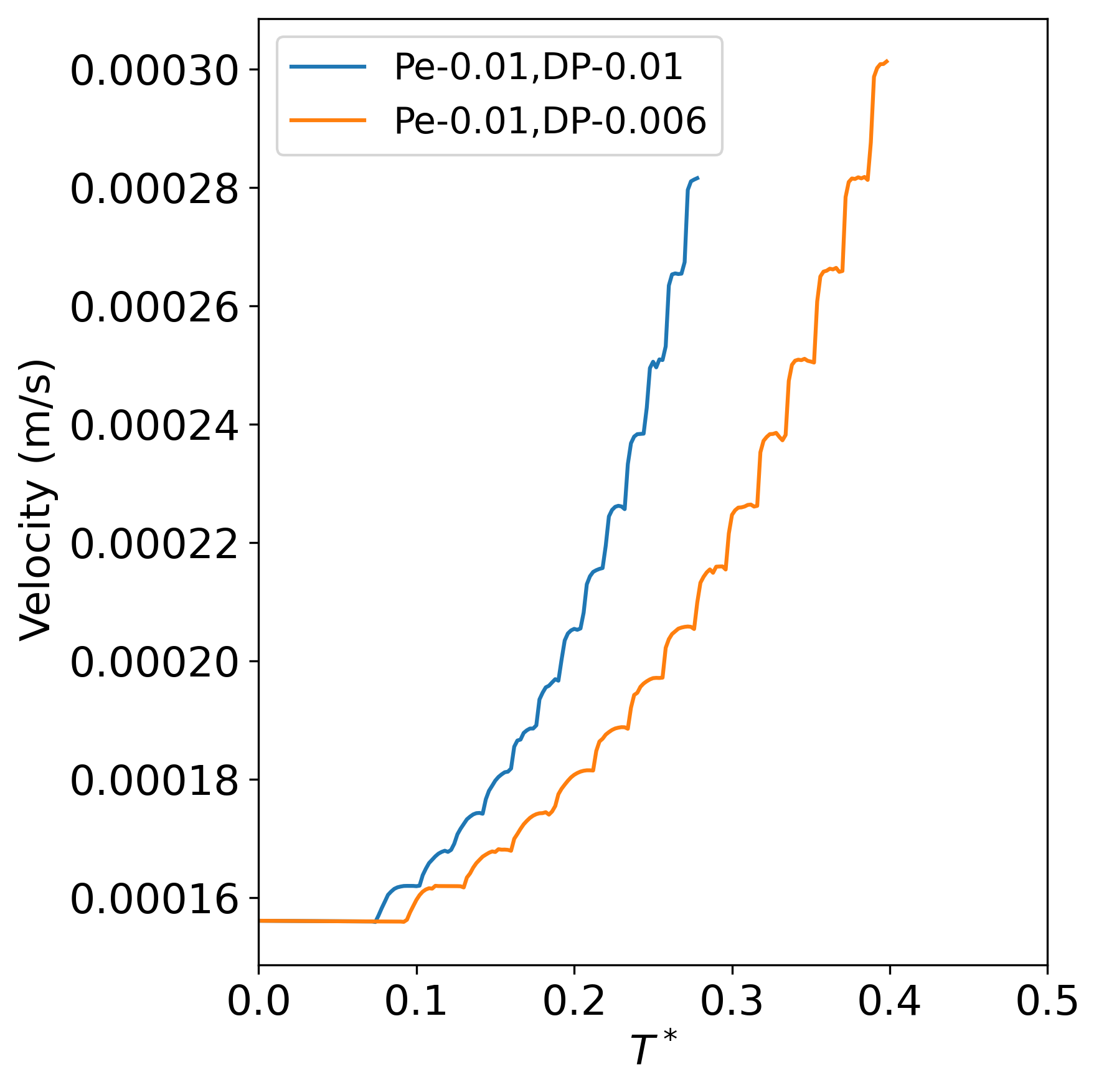} 
        (b)
    \end{minipage}
        \begin{minipage}[b]{0.45\textwidth}
        \centering
        \includegraphics[clip, trim=0 0 0 0, width=\textwidth]{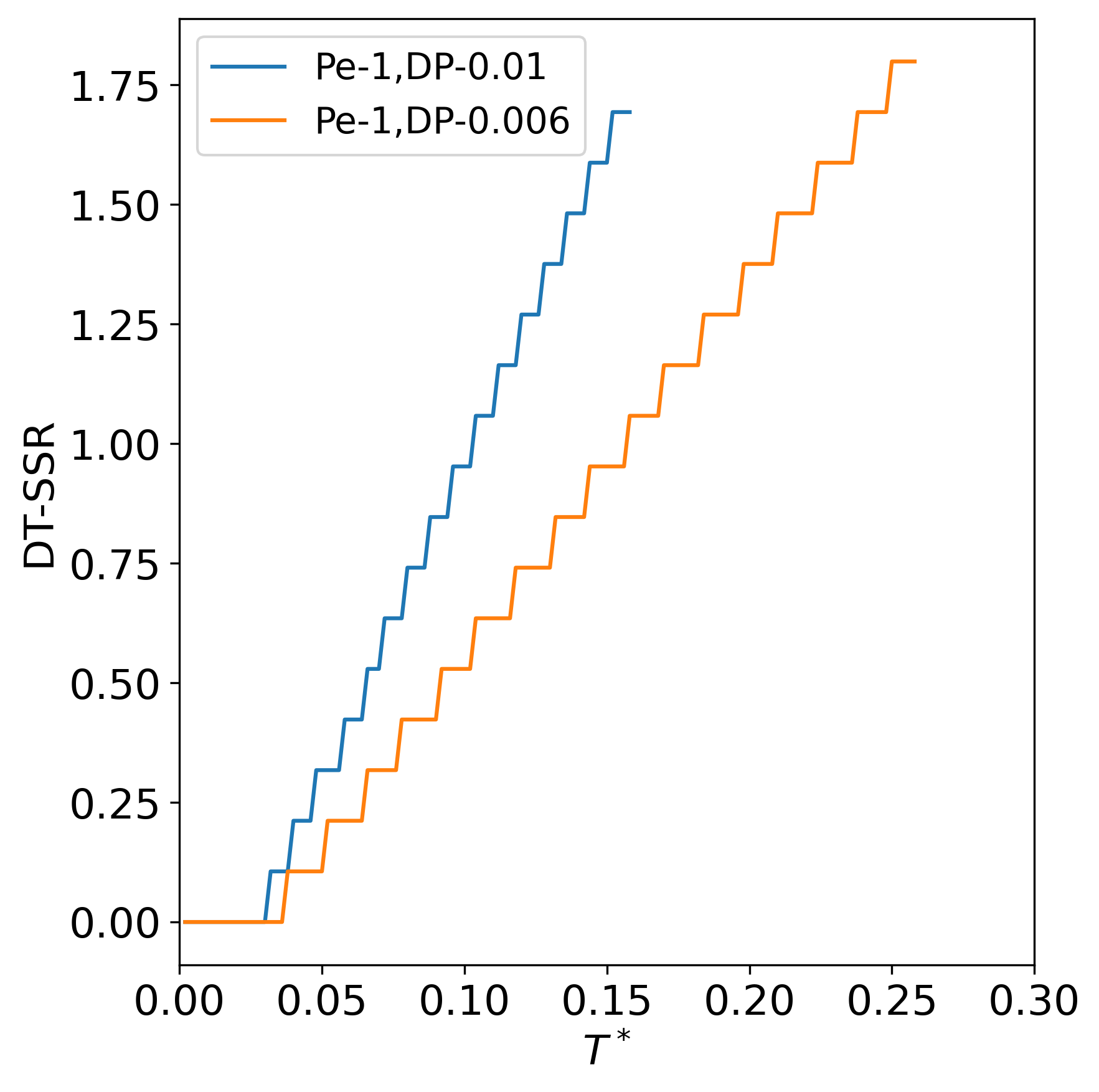} 
        (c)
    \end{minipage}
    \hfill 
    \begin{minipage}[b]{0.45\textwidth}
        \centering
        \includegraphics[width=\textwidth]{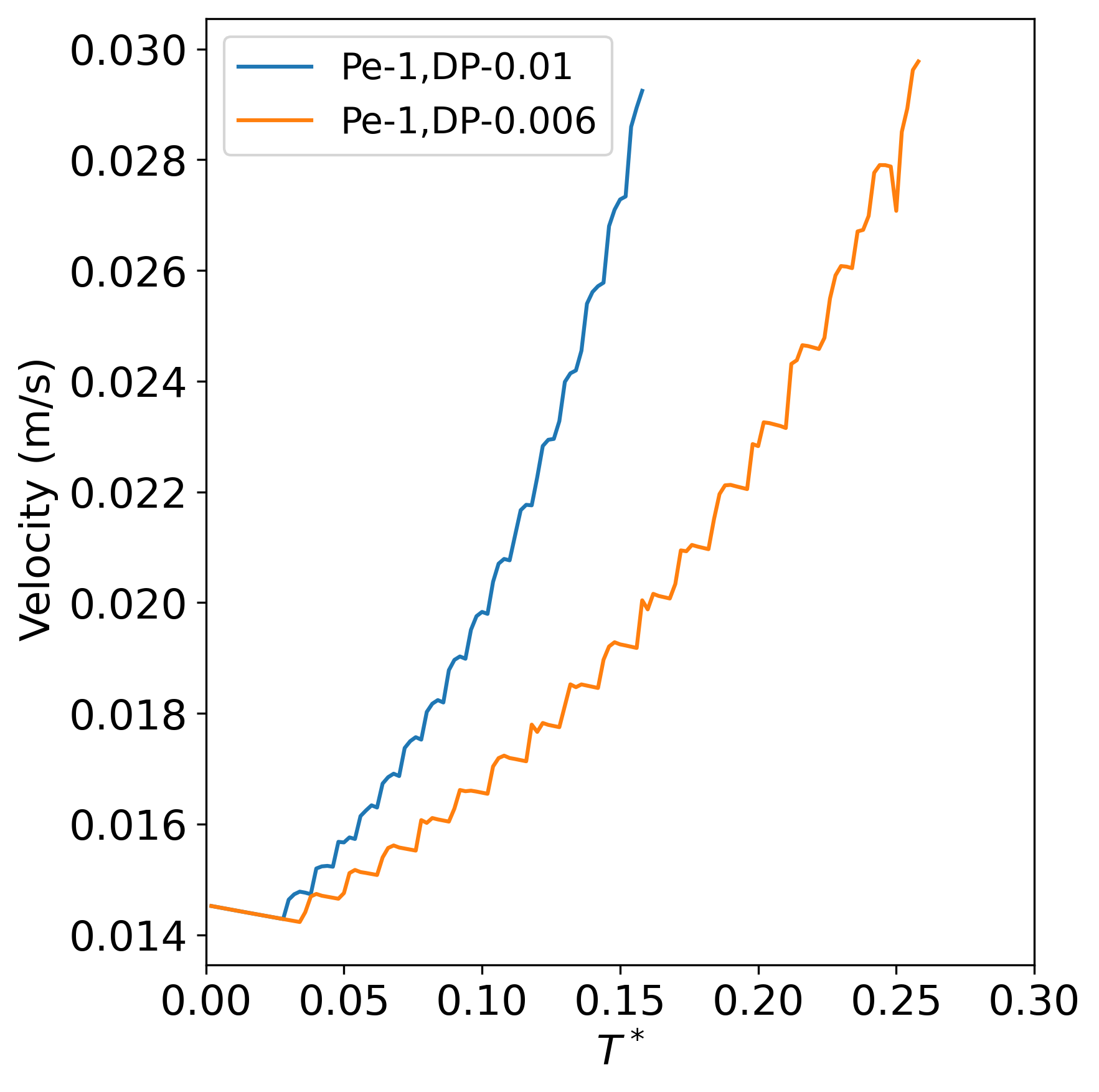} 
        (d)
    \end{minipage}
     \caption{The evolution of particle deposition around the substrate and the maximum velocity component in the fluid domain over time is given. (a) The deposition ratio and (b) the maximum velocity component in the fluid domain are given over time for $Pe = 0.01$ and inlet mass fraction of concentration $Y_c =0.01$, comparing two cases with particle deposition probabilities $DP=0.006$ and $0.01$. (c) The deposition ratio and (d) the maximum velocity component in the fluid domain are shown over time for $Pe = 1$, comparing two cases with particle deposition probabilities $DP=0.006$ and $0.01$.}
    \label{fig:result2}
\end{figure}

  Furthermore, the quantitative relationship between particle deposition and varying $Pe$ and $DP$ is presented in Figure \ref{fig:result2}. Deposition is quantified in terms of the DT-SSR, defined as the ratio of deposition thickness to substrate size. Firstly,  the deposition ratio of particles are presented onto the substrate with changing $DP$. Secondly, $Pe$ is varied from 0.01 to 1 by changing the average inlet velocity prior to particle deposition while keeping other parameters constant. As shown in Figure \ref{fig:result2}(a), the deposition ratio increases at a faster rate at $DP = 0.01$ compared to $DP = 0.006$, as explained the cause above paragraph. The effect of this faster particle deposition is visualized through the increase in the maximum x-component velocity in the fluid with time, as shown in Figure \ref{fig:result2}(b). This can be explained through the conservation of the fluids mass. As particle deposition increases, the number of fluid nodes decreases. However, since the fluid is incompressible, the velocity must increase to conserve mass in the system.  
Moreover, the particle deposition occurs much faster rate with increase in simulation time for $DP = 0.01$ than $DP = 0.006$. This can be attributed to the fact that, as deposition happens more rapidly at $DP = 0.01$, more particles are recruited toward the substrate. This therefore increases contact with additional particles, causing the deposition thickness to widen. Additionally, DP significantly influences deposition; a twofold increase in deposition probability accelerates particle deposition by approximately twice as fast.  A similar trend in the change of maximum x-component velocity and DT-SSR is observed, as shown in Figure \ref{fig:result2}(c), with corresponding maximum velocity change over time presented in Figure \ref{fig:result2}(d).

  Thirdly, we compare the particle deposition behavior between $Pe= 0.01$ and $1$. The results show that as $Pe$ increases, the thickness of the particle deposition on the substrate grows more rapidly over time, as compared Figure \ref{fig:result2}(c) with  Figure \ref{fig:result2}(c) for all $DP$. This leads to a longer time lag for particle deposition on the substrate at $Pe = 0.01$ compared to $Pe = 1$ before particles start accumulating. The faster rate of particle deposition at $Pe = 1$ is due to the higher imposed flow velocity, which convects more solutes toward the substrate, where they are eventually adsorbed and converted to the solid substrate.
\subsubsection{Effect of the inlet mass fraction of particles}
\begin{figure}[!hbt]
    \begin{minipage}[b]{0.5\textwidth}
        \centering
        \includegraphics[clip, trim=0 0 0 0, width=\textwidth]{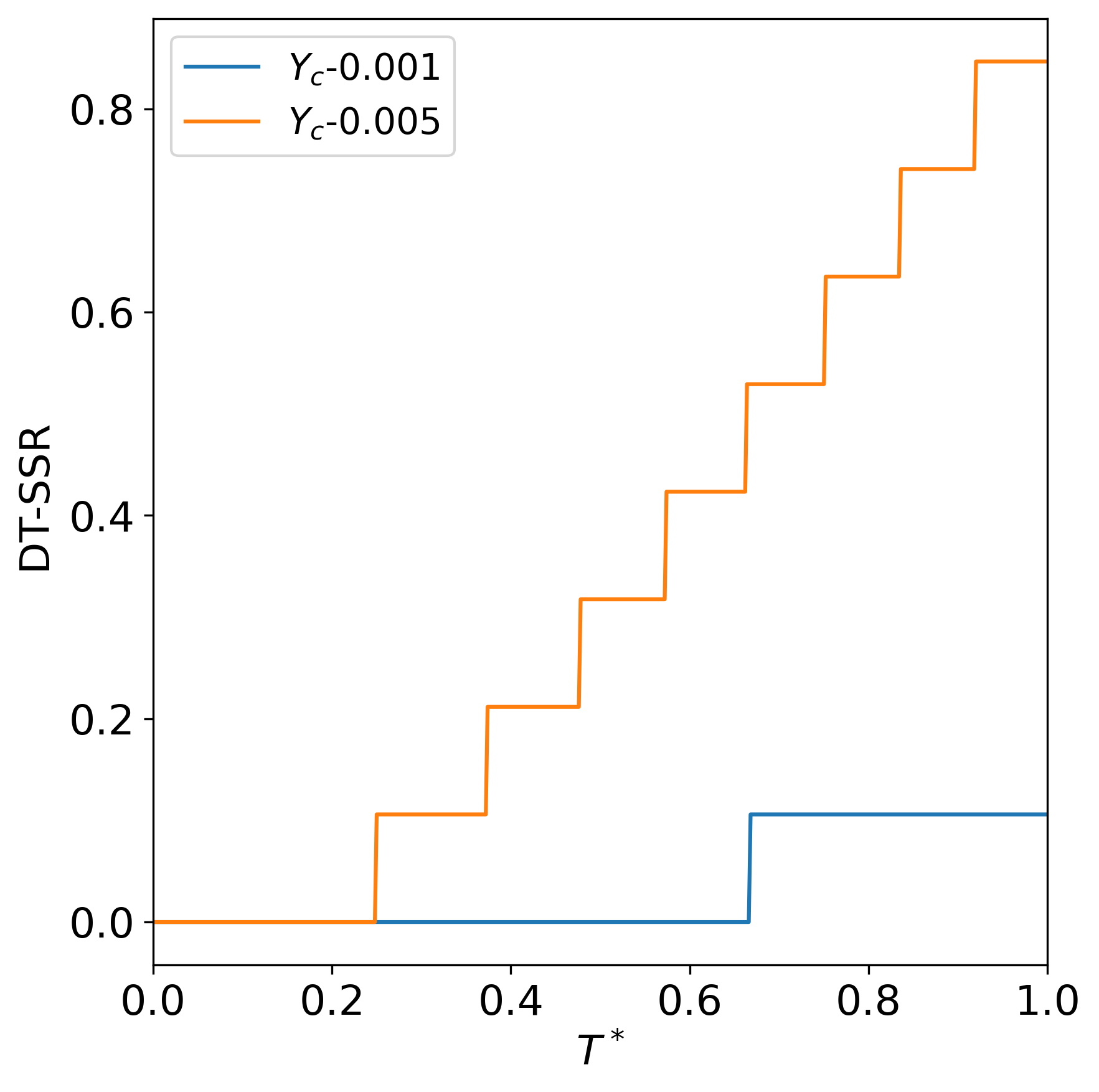} 
        (a)
    \end{minipage}
    \hfill 
    \begin{minipage}[b]{0.5\textwidth}
        \centering
        \includegraphics[width=\textwidth]{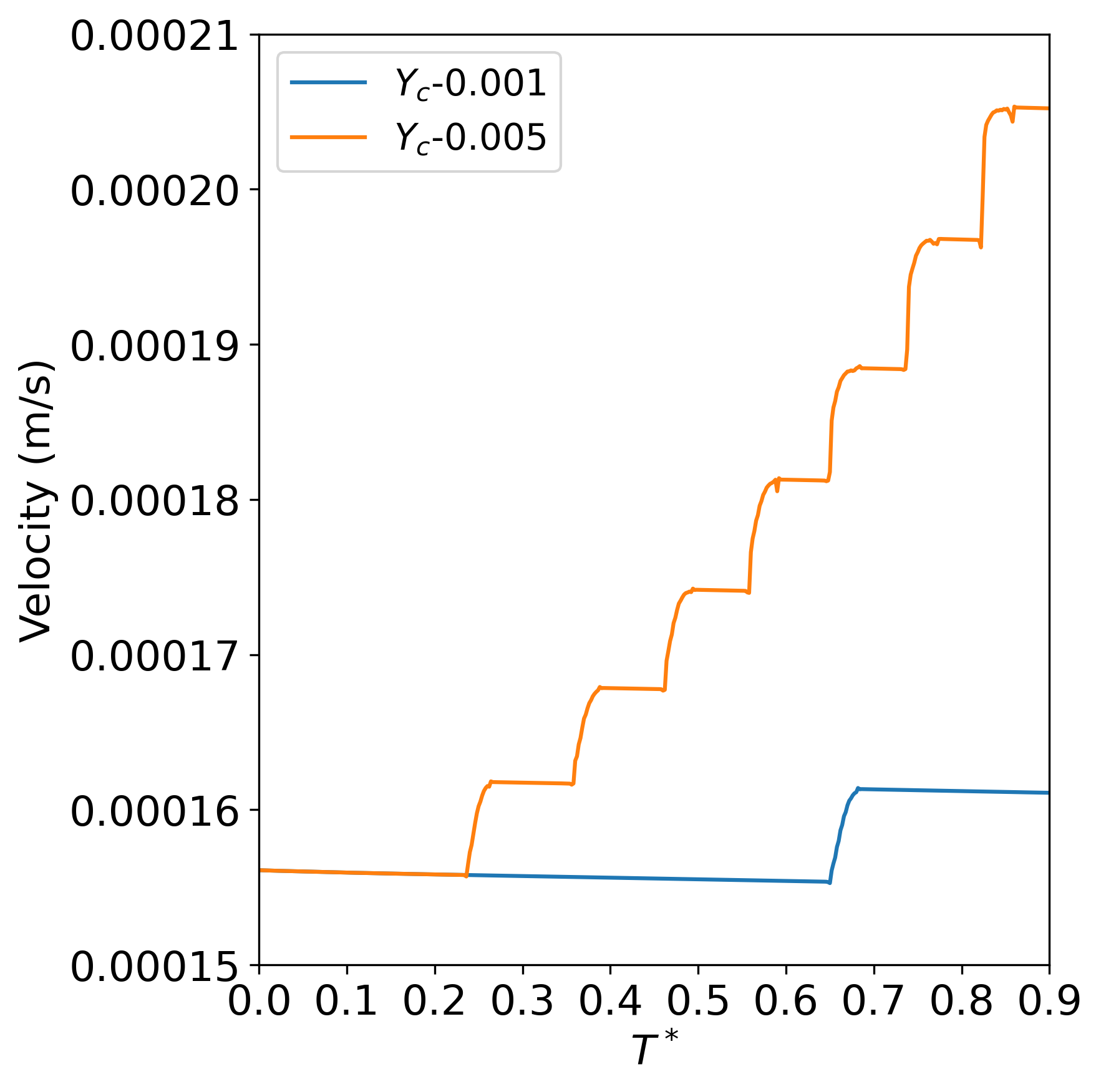} 
        (b)
    \end{minipage}
     \caption{The evolution of particle deposition around the substrate and the maximum velocity component in the fluid domain over time at $Pe = 0.01$ and $PD = 0.002$ is presented.  (a) The deposition ratio and (b) the maximum velocity component in the fluid domain are shown over time, comparing two cases inlet mass fractions of the concentration field $Y_c = 0.005$ and 0.001.}
    \label{fig:result3}
\end{figure}

The deposition of particles is explicitly connected to the mass fraction of concentration in the vicinity of the substrate (given in equation \eqref{surface_deposition}). To understand how changes in the inlet mass fraction of particles influence deposition on the substrate, we consider two inlet concentrations i.e., 0.005 and 0.001, at $Pe = 0.01$ and $PD = 0.002$.  Figure \ref{fig:result3} presents the deposition ratio on the substrate and the maximum x-component of velocity over time, by varying the inlet mass fraction of particles between 0.005 and 0.001. It is evident from Figure \ref{fig:result3}(a) that with an increasing inlet mass fraction of particles, deposition occurs at a much faster rate. This deposition does not occur immediately on the substrate surface. Initially, the surface mass fraction $Y_{c,s}$ begins to accumulate before being deposited. There is a time gap between each subsequent deposition, evidenced by a lack of change in DT-SSR values, indicated by horizontal lines as shown in Figure \ref{fig:result3}(a). This gap is prolonged with a smaller inlet mass fraction, such as 0.001. Furthermore, the higher deposition causes an increase in the maximum x-component velocity due to the conservation of mass in the fluid domain, as shown in Figure \ref{fig:result3}(b).

\subsection{Particle deposition on multiple substrates }
The investigation in the previous section centered on particle deposition onto a single substrate. However, the deposition behavior of the particles when multiple substrates are positioned in tandem within the flow remains uncertain. In this study, the effect of particle deposition on downstream substrates is evaluated. Four substrates, designated O1, O2, O3, and O4, are arranged sequentially from upstream to downstream, aligned horizontally with the flow direction inside the channel. The axial center of substrate O1 is located 110 lattice nodes away from the inlet, with a gap of 20 lattice nodes between substrates. In this section,  the accumulation and subsequent deposition of particles on the substrate surfaces by changing the same parameters as in the previous section is analyzed. The main objective of this section is to present the differences in particle deposition between upstream and downstream substrates.
\subsubsection{Effect of flow velocity and deposition probability}
\begin{figure}[!hbt]
\centering 
    \begin{minipage}[b]{0.8\textwidth}
        \centering
        \includegraphics[clip, trim=0 0 0 0, width=\textwidth]{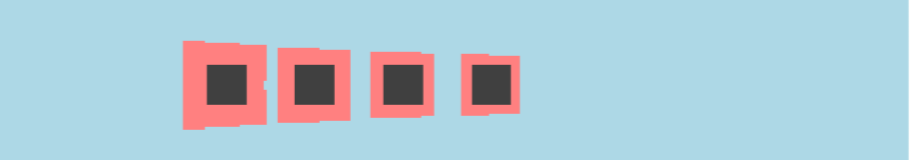} 
        (a)
    \end{minipage}
   \vspace{0.5cm} 
    \begin{minipage}[b]{0.8\textwidth}
        \centering
        \includegraphics[width=\textwidth]{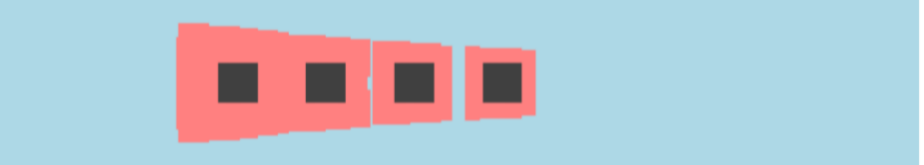} 
        (b)
    \end{minipage}
     \caption{Simulation snapshots show the particle deposition around the substrates at time $T^\star$ = 0.2  for given $Pe = 0.01$ and inlet particle mass fraction $Y_c =0.01$, comparing two cases of particle deposition probability ($DP$): (a) 0.006 and (b) 0.01.}
    \label{fig:result4}
\end{figure}

Figure \ref{fig:result4}, for $Pe = 0.01$ and $DP = 0.006$ and $0.01$ at time $T^\star = 0.2$, shows the deposition pattern across the substrates. From that figure, the deposition pattern demonstrates that particle deposition on the upstream substrate (O1) is higher than that on the downstream substrate (O4). This is due to the fact that, at the beginning of the simulation, substrate O1 is the first to be exposed to solute particles flowing from the inlet of the channel, leading to greater deposition.  As deposition progresses around the substrate O1, it affects the flow field around the substrate. Consequently, the streamlines of fluid carrying solute particles are eventually altered, affecting the transport of solute particles to the downstream substrates. Furthermore, this pattern does not change even with variations in $DP$. However, the overall rate of deposition across all substrates is more at $DP = 0.01$, Figure \ref{fig:result4}(b), compared to $DP = 0.006$ as shown in Figure \ref{fig:result4}(a). This indicates that a higher $DP$ leads to increased deposition of the particles, resulting in faster growth of the deposition layer around the substrate.

\begin{figure}[!hbt]
    \begin{minipage}[b]{0.5\textwidth}
        \centering
        \includegraphics[clip, trim=0 0 0 0, width=\textwidth]{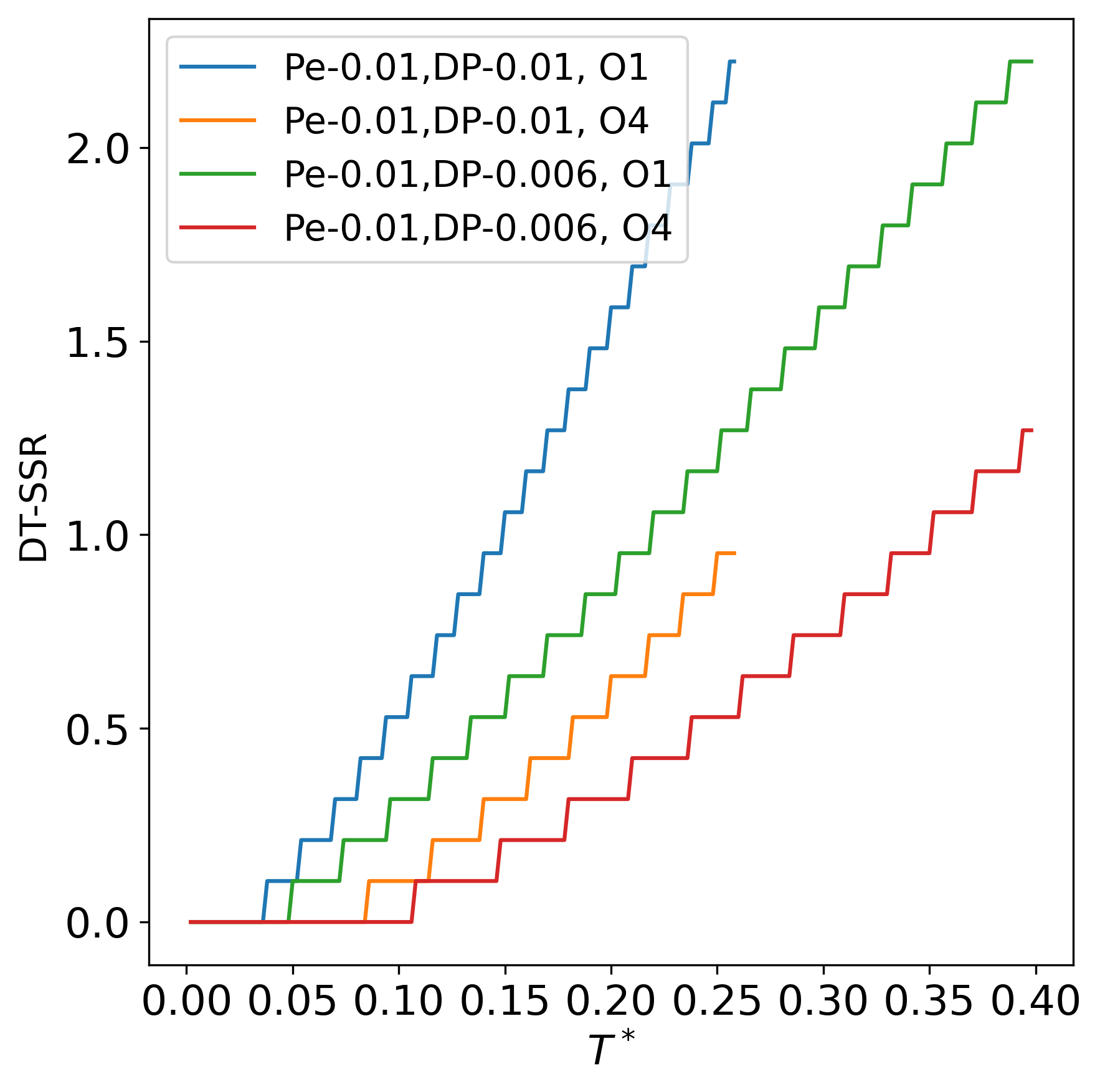} 
        (a)
    \end{minipage}
    \hfill 
    \begin{minipage}[b]{0.5\textwidth}
        \centering
        \includegraphics[width=\textwidth]{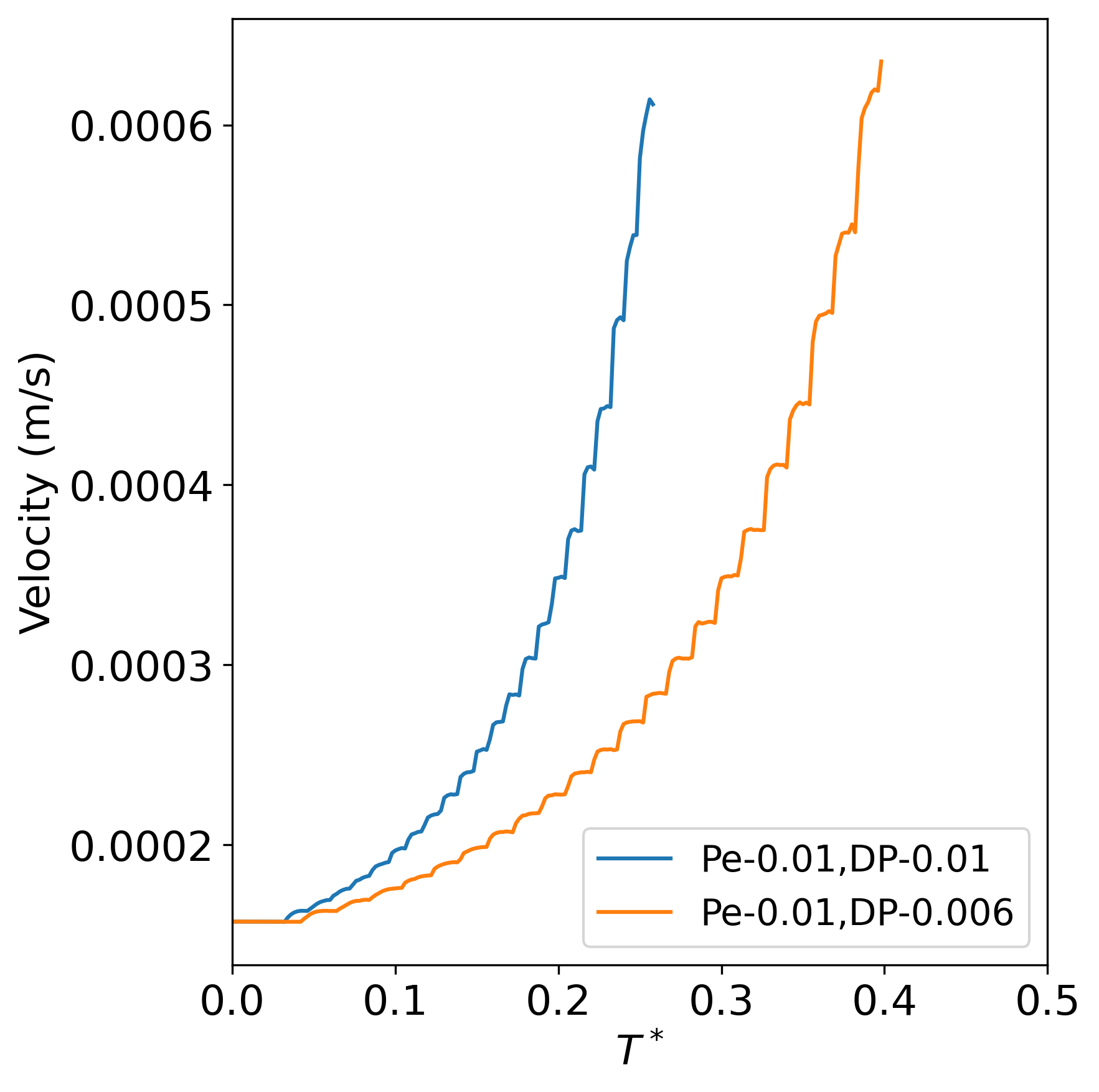} 
        (b)
    \end{minipage}
        \begin{minipage}[b]{0.5\textwidth}
        \centering
        \includegraphics[clip, trim=0 0 0 0, width=\textwidth]{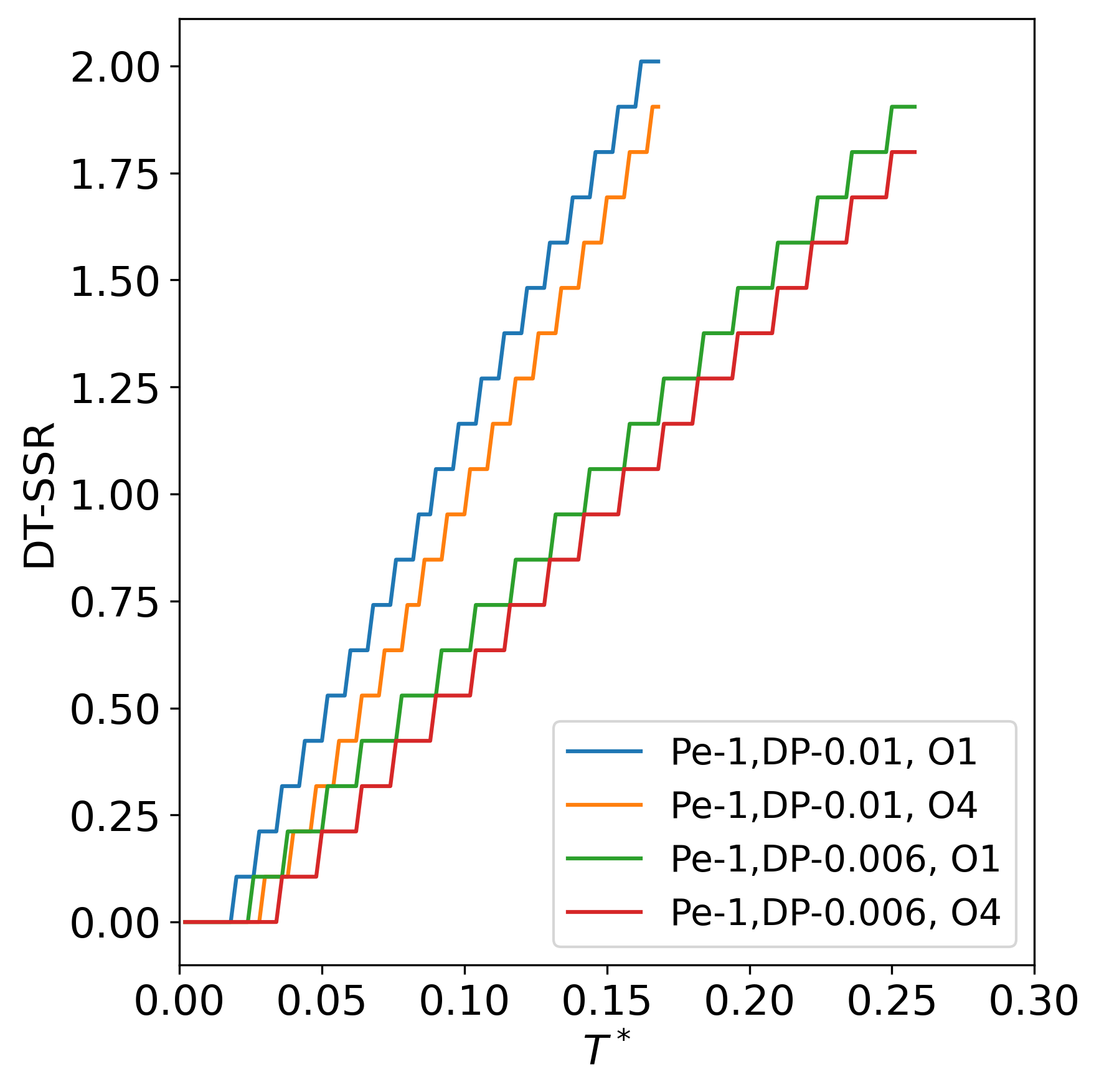} 
        (c)
    \end{minipage}
    \hfill 
    \begin{minipage}[b]{0.5\textwidth}
        \centering
        \includegraphics[width=\textwidth]{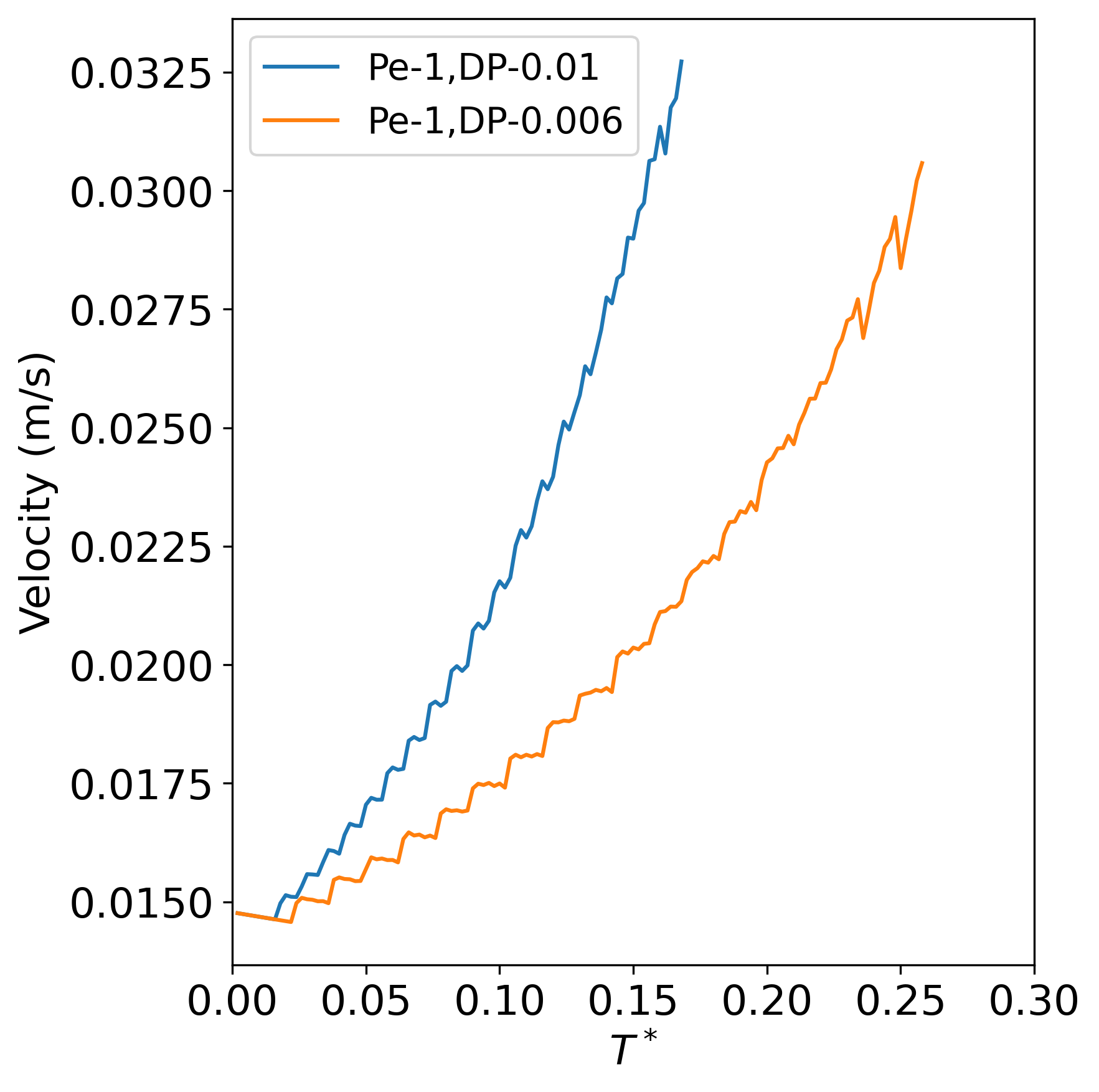} 
        (d)
    \end{minipage}
     \caption{The evolution of particle deposition around the substrates—namely, the upstream substrate (O1) and the downstream substrate (O4)—and the maximum velocity component in the fluid domain over time are presented. (a) The deposition ratio and (b) the maximum velocity component in the fluid domain are shown with time for $Pe = 0.01$ and inlet mass fraction of concentration $Y_c = 0.01$, comparing two cases with particle deposition probabilities $DP = 0.006$ and 0.01. (c) The deposition ratio and (d) the maximum velocity component in the fluid domain are presented with time for $Pe = 1$ and inlet mass fraction of concentration $Y_c = 0.01$, comparing two cases with particle deposition probabilities $DP = 0.006$ and $0.01$.}
    \label{fig:result5}
\end{figure}

  In order to provide a quantitative overview of particle deposition and velocity variation in the channel over time, we present the DT-SSR around substrates O1 and O4 by varying $Pe$ and $DP$, as shown in Figure \ref{fig:result5}. Figure \ref{fig:result5}(a) shows the particle deposition ratio over time for $Pe = 0.01$, with $DP = 0.006$ and 0.01. The deposition ratios for both cases are compared between substrates O1 and O4. As discussed earlier, the particle deposition occurs at a faster rate for $DP = 0.01$ on both O1 and O4 than $DP = 0.006$. Consequently, the magnitude of particle deposition is higher on O1 and O4 for $DP = 0.01$. The initial delay before particles begin to deposit on substrates O1 and O4 is longer at $DP = 0.006$ than for $DP = 0.01$. However, the subsequent deposition takes slightly longer time on substrate O4 at $DP = 0.006$ than at $DP = 0.01$. Additionally, due to the higher deposition rate, the velocity must increase to maintain conservation of mass in the fluid. Therefore, Figure \ref{fig:result5}(b) shows that the maximum velocity is higher at $DP=0.01$ than at $DP = 0.006$.

  Similarly, the deposition of particles and velocity variation by setting $Pe = 1$ for $DP = 0.006$ and 0.01 are analyzed. Figure \ref{fig:result5}(c) shows the dynamic changes in particle deposition for $Pe = 1$ at $DP = 0.006$ and $0.01$. Two interesting observations are noted: (a) the deposition of particles on O1 and O4 occurs at a similar rate for both $DP = 0.006$ and $DP = 0.01$, and (b) the subsequent deposition rates, before lattice nodes are converted to solids, are similar in both cases. However, the deposition magnitude on both O1 and O4 at $DP = 0.01$ is consistently higher than that of $DP = 0.006$. The slopes of deposition growth on the substrates are linear and are highly dependent on $DP$. When compared to $Pe = 0.01$, shown in Figure \ref{fig:result5}(a), the initial and subsequent deposition rates are lower on all substrates at $Pe = 0.01$ than at $Pe = 1$, as shown in Figure \ref{fig:result5}(c). This occurs because a higher $Pe$ corresponds to an increased velocity, which transports solute particles more rapidly and initiates more uniform deposition across all substrates. In contrast, at lower $Pe$, deposition rates vary between substrates, as shown in Figure \ref{fig:result5}(a). Additionally, for $Pe = 1$, deposition at $DP = 0.01$ on all substrates consistently remains higher than at $DP = 0.006$, as shown in Figure \ref{fig:result5}(c). 

  However, this condition does not hold at $Pe = 0.01$. For $Pe = 0.01$, deposition on substrate O4 at both $DP = 0.006$ and $DP = 0.01$ is always lower than on substrate O1 at the same $DP$, as shown in Figure \ref{fig:result5}(a). Moreover, the accumulation of particles on the substrates correlates with significant changes in fluid dynamics. As particle deposition increases, the effective flow area decreases, leading to a noticeable increase in maximum flow velocity, as shown in Figure \ref{fig:result5}(d).

\subsubsection{Effect of the inlet mass fraction of particles}
\begin{figure}[!hbt]
    \begin{minipage}[b]{0.5\textwidth}
        \centering
        \includegraphics[clip, trim=0 0 0 0, width=\textwidth]{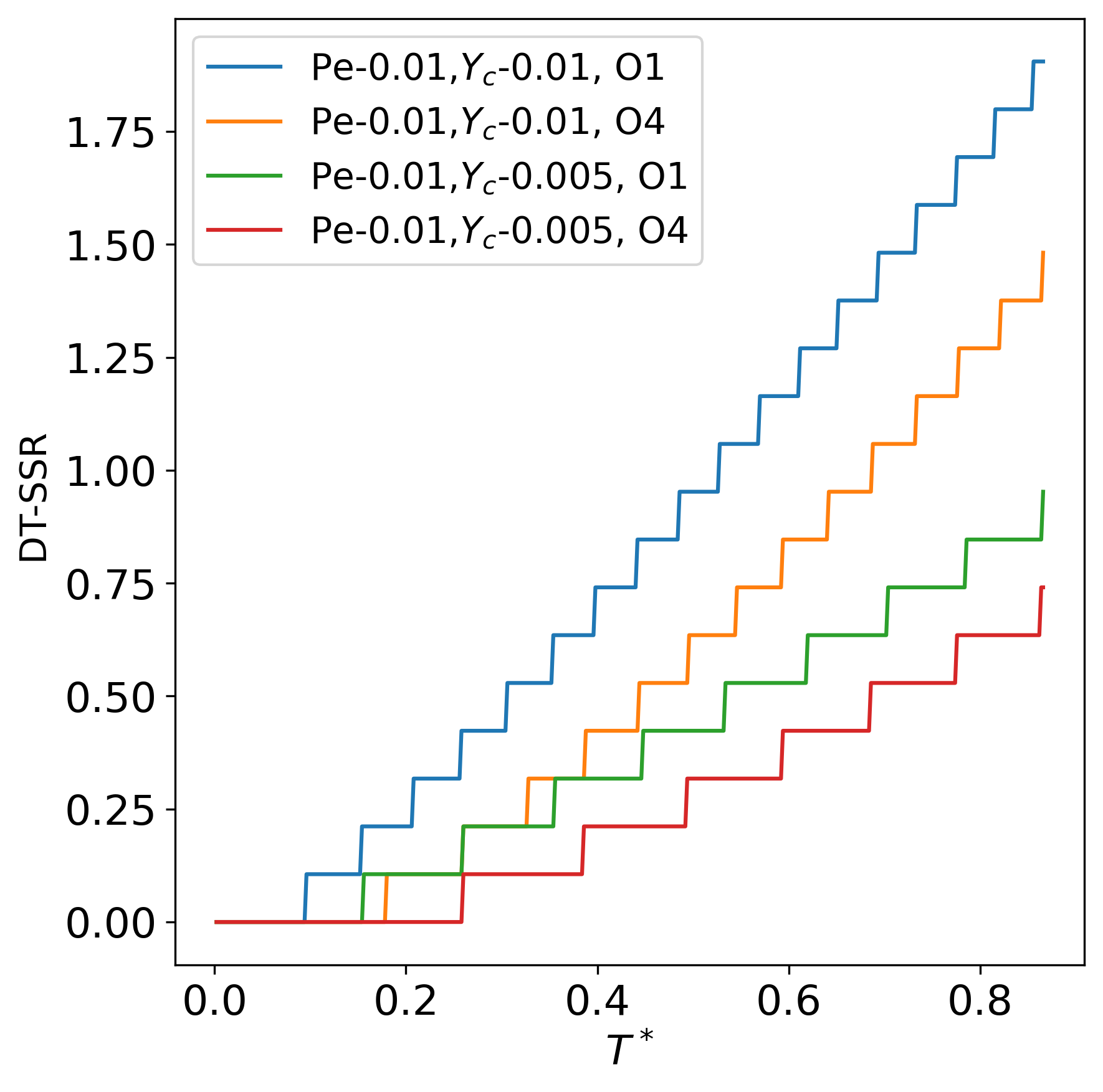} 
        (a)
    \end{minipage}
    \hfill 
    \begin{minipage}[b]{0.5\textwidth}
        \centering
        \includegraphics[width=\textwidth]{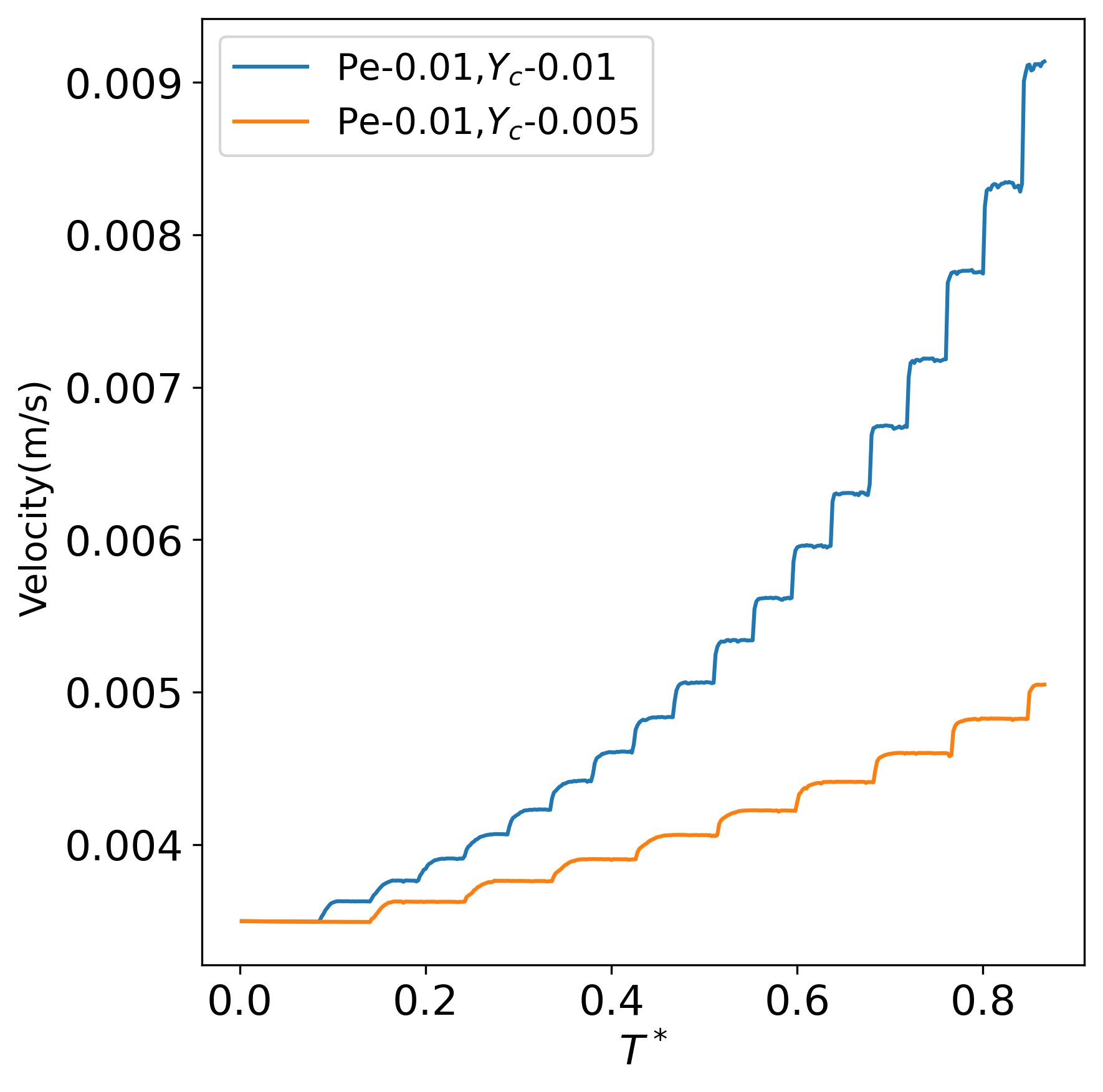} 
        (b)
    \end{minipage}
 \caption{The evolution of particle deposition around the substrates—specifically, the upstream substrate (O1) and the downstream substrate (O4)—as well as the maximum velocity component in the fluid domain, are presented over time for $Pe = 0.01$ and a particle deposition probability of $DP = 0.002$. (a) The deposition ratio and (b) the maximum velocity component in the fluid domain are shown over time, comparing two cases of varying inlet particle mass fraction $Y_c = 0.005$ and 0.001.}
    \label{fig:result6}
\end{figure}

In this section, we investigate how variations in the inlet particle mass fraction affect particle deposition around the substrates, using the same configuration as presented in the previous section. Specifically, we examine the impact of particle mass fractions on deposition at $Pe = 0.01$ and $DP = 0.002$, with inlet particle mass fractions set to 0.005 and 0.01. Figure \ref{fig:result6}(a) demonstrates the effect of mass concentration variation on deposition of the particles at substrates O1 and O4. At an inlet particle mass fraction of 0.01, the deposition rate on substrate O4 is significantly lower than on substrate O1. This trend persists at lower mass fractions, such as 0.005, where the overall deposition is reduced relative to that at 0.01.

  This outcome can be attributed to the fact that increasing the inlet particle mass fraction enhances particle transport toward the substrates, leading to greater accumulation on the substrate surface, which solidifies as deposited particles at fluid nodes. Additionally, lower inlet mass fractions require more time to initiate deposition, with subsequent deposition rates progressing more slowly. The higher deposition of particles around the substrates at an inlet mass fraction of 0.01 also affects the velocity profile, as shown in Figure \ref{fig:result6}(b), where the maximum velocity is higher at an inlet particle mass fraction of 0.01 in contrast to 0.005.
\subsection{Particle deposition on regularly patterned substrates}
\begin{figure}[!hbt]
    \centering 
    \begin{minipage}[b]{0.8\textwidth} 
        \centering
        \includegraphics[clip, trim=0 0.0 0 0, width=\textwidth]{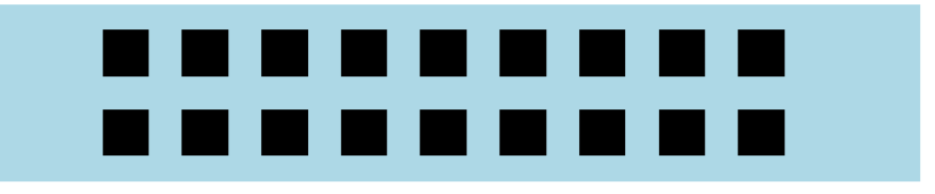} 
        (a)
    \end{minipage}

    \vspace{0.5cm} 
    \begin{minipage}[b]{0.8\textwidth} 
        \centering
        \includegraphics[clip, trim=0 0.0 0 0, width=\textwidth]{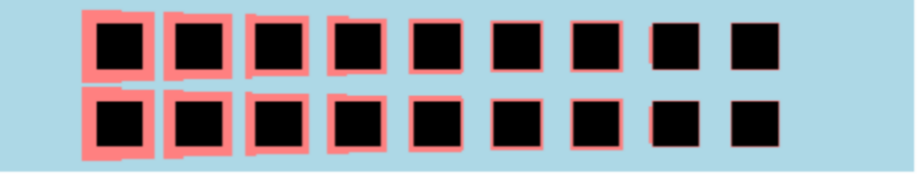} 
        (b)
    \end{minipage}
    \caption{Simulation snapshots show particle deposition around the regular patterned substrates at $Pe = 0.01$, particle deposition probability $DP = 0.002$, and inlet particle mass fraction $Y_c = 0.01$: (a) $T^\star = 0$ and (b) $T^\star = 0.27$.}
    \label{fig:result7}
\end{figure}

In actual porous media, the structure consists of a complex network of pores where particle deposition occurs \cite{yamamoto2009simulation}. In our study, we considered a simplified, regularly patterned array of square substrates arranged in two rows, with each row containing 9 substrates, as shown in Figure \ref{fig:result7}(a). The horizontal and vertical spacing between the substrates is 16 lattice nodes. The centers of the substrates in both the upper and lower rows are equidistant from the upper and lower channel walls, at a distance of 22 lattice nodes. Additionally, the centers of the first and last columns are situated 60 and 72 lattice nodes away from the inlet and outlet of the channel, respectively.

  Figure \ref{fig:result7}(b) shows the particle deposition on different substrates at $Pe = 0.01$. The deposition of particles is more asymmetric than that observed in the case with four substrates (Figure \ref{fig:result4}). The deposition pattern indicates that the first substrate exhibits a higher concentration of particle deposition than the subsequent substrates. This variation is attributed to the increased interaction of particles at the beginning of the channel, where a higher concentration of particles initially contacts the first column. The substrates in the last column take a longer time to accumulate deposits relative to the upstream substrates.

\begin{figure}[!hbt]
        \centering 
    \begin{minipage}[b]{0.8\textwidth} 
        \centering
        \includegraphics[clip, trim=0 0.0 0 0, width=\textwidth]{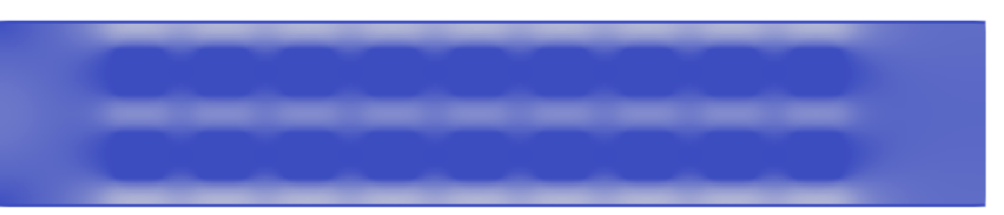} 
        (a)
    \end{minipage}
    \vspace{0.5cm} 

        \centering 
    \begin{minipage}[b]{0.8\textwidth} 
        \centering
        \includegraphics[clip, trim=0 0.0 0 0, width=\textwidth]{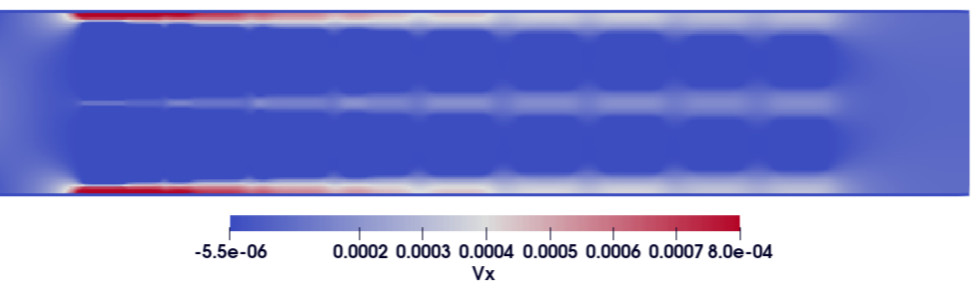} 
        (b)
    \end{minipage}
    \caption{Simulation results show the velocity component along the flow direction at $Pe = 0.01$, particle deposition probability $DP = 0.002$, and inlet particle mass fraction $Y_c = 0.01$: (a) $T^\star = 0$ and (b) $T^\star = 0.27$.}
    \label{fig:result8}
\end{figure}

  An interesting observation is that the center between the two rows becomes blocked at the upstream substrates due to deposition, unlike the gaps between the walls and the substrates, which remain less affected. The effects of particle deposition are evident in the velocity field snapshots of the channel, as shown in Figure \ref{fig:result8}. As deposition increases, there is a noticeable change in velocity around the initial few columns of substrates, as illustrated in Figure \ref{fig:result8}(b).

\subsection{Particle deposition in an interconnected porous structure}
In practical industrial applications, porous structures often exhibit considerable complexity. For example, diesel particulate filters (DPFs) deploy a honeycomb-like porous architecture to capture PM. An in-depth understanding of the particle deposition processes within these structures is essential for optimizing their design and enhancing overall performance. In this section, we simulate our deposition model using a 2-D geometry obtained through Focused Ion Beam Scanning Electron Microscopy (FIB-SEM). 
\begin{figure}[!hbt]
        \centering 
        \includegraphics[clip, trim=0 0.0 0 0, width=\textwidth]{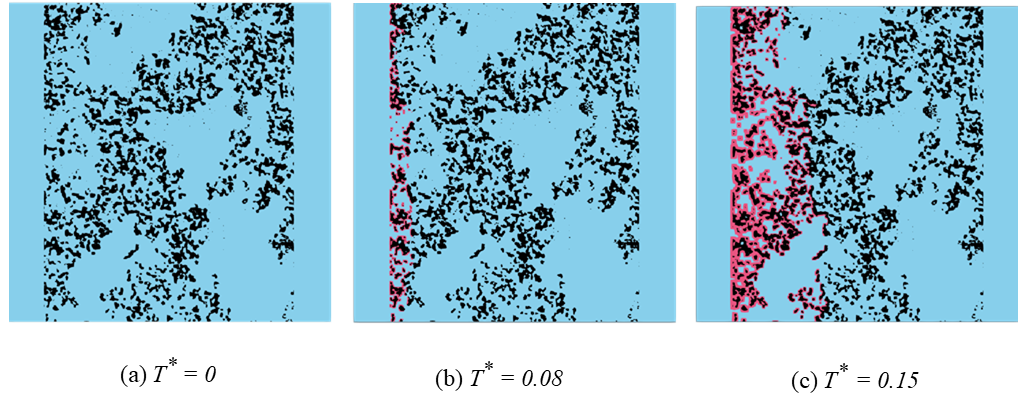} 
    \caption{Simulation snapshots illustrate the deposition of particles over a complex porous structure under the conditions: $Pe = 0.01$, particle deposition probability $DP = 0.002$, and inlet particle mass fraction $Y_c = 0.01$. The snapshots correspond to (a) $T^\star = 0$, (b) $T^\star = 0.08$, and (c) $T^\star = 0.15$. The different colors indicate: deposited particles (red), fluid (sky blue), and substrate (black).}
    \label{fig:result9}
\end{figure}
The size of the simulated domain is $L_1 \times W_1 = 419 \times 415$, with a steady parabolic velocity profile at the inlet. The maximum input velocity, $v_{\text{max}}$, is 5.65e-6, and the characteristic length scale is $W_1=415$, resulting in a Péclet number ($Pe$) of 0.01. Additionally, the deposition probability ($DP$) and inlet particle mass fraction ($Yc$) are 0.002 and 0.01, respectively. The remaining parameters are identical to those used in the simulation with square obstacles.
\begin{figure}[!hbt]
        \centering 
        \includegraphics[clip, trim=0 0.0 0 0, width=\linewidth]{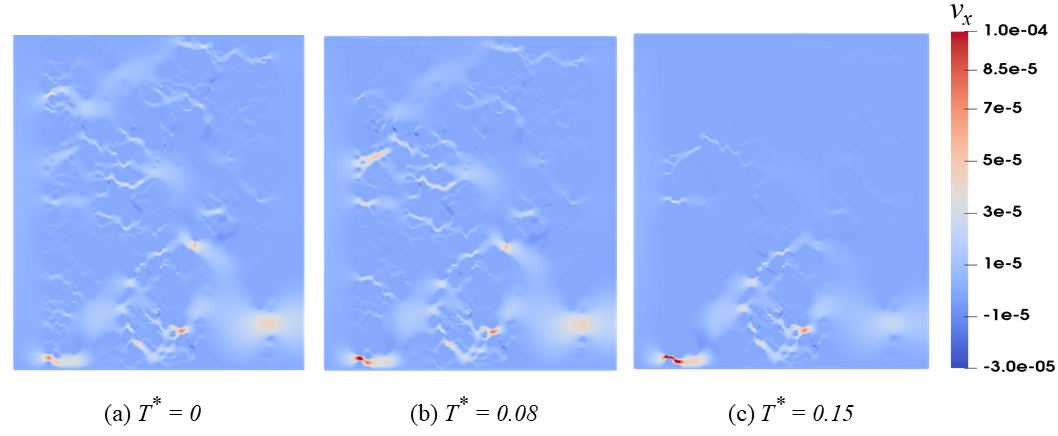} 
    \caption{Simulation snapshots show the x-component velocity under the conditions: $Pe = 0.01$, particle deposition probability $DP = 0.002$, and inlet particle mass fraction $Y_c = 0.01$. The snapshots correspond to (a) $T^\star = 0$, (b) $T^\star = 0.08$, and (c) $T^\star = 0.15$.}
    \label{fig:result10}
\end{figure}

  Figure \ref{fig:result9} illustrates deposition of the particles within the simulated domain at various time steps. These figures clearly demonstrate that the pore structure is randomly distributed across the domain. We deliberately added 50 lattice nodes at both the inlet and outlet, as depicted in Figure \ref{fig:result9}(a). This approach stabilizes the simulation and enables a detailed analysis of the particle deposition patterns around randomly distributed obstacles. As particles cross the obstacle-laden domain, they become trapped in smaller pores near the inlet, where initial interactions with obstacles occur. This phenomenon is highlighted by the red regions in Figure \ref{fig:result9}(b). As time progresses, the deposition profile extends gradually downstream through the porous network. At the same time, particle accumulation increases further near the inlet, eventually restricting fluid flow. This process results in two important consequences: (a) increased blockage near the inlet, affecting particles transport to the the downstream, and (b) limited particle deposition further downstream, preventing significant accumulation near the outlet, as shown in Figure \ref{fig:result9}(c). To avoid reporting incorrect results, we terminated the simulation at $T\star = 0.15$. Additionally, the results presented here are consistent with those reported in the previous sections.

Figure \ref{fig:result10} presents the x-component velocity $v_x$ for all reported cases, as shown in Figure \ref{fig:result9}. Initially, the simulation domain was initialized with a steady velocity profile, without solving for particle deposition, as depicted in Figure \ref{fig:result10}(a). Distinct narrow white and red bands are observed, representing the fluid flow pathways and indicating the pores between the obstacles. As particle deposition progresses, as illustrated in Figure \ref{fig:result10}(b), the pore diameter decreases, leading to an increase in velocity to maintain fluid continuity. Further deposition, as shown in Figure \ref{fig:result10}(c), results in partial or complete blockage of the fluid flow, which causes a reduction in narrow white band and an increase in velocity through the pores where minimal particle deposition occurs.
\section{Conclusion}
\label{conclusion}
In this work, a detailed study of particle deposition on substrates is conducted using an in-house developed LBM code. The LBM algorithm for solving the coupled continuity, momentum, and advection-diffusion equations is thoroughly elaborated. Particle deposition is evaluated using key metrics such as deposition probability, mass concentration of particles in the fluid, and the velocity of the flow. We have considered various test cases, including single, multiple, and regularly arranged square shaped substrate configurations as well as complex interconnected porous network in a two-dimensional model.
The deposition of particles on the substrate is quantified by calculating the DT-SSR. It is found that the DT-SSR is significantly influenced by flow rate, deposition probability, particle mass fraction, the number of substrates, and their positions. This simplified two-dimensional model will offer valuable insights into more complex particle deposition systems and enhance our understanding of deposition dynamics in real-world applications.
\section*{Acknowledgments}
This work is supported by the “CARNOT ESP” program, through the project “HYBRID”. The authors gratefully acknowledge the access to French HPC resources provided by the French regional computing center of Normandy CRIANN (2022006). 
\section*{Declarations}
\begin{itemize}
\item \textbf{Conflict of interest/Competing interests:} The authors have no financial or proprietary interests in any material discussed in this article.
\item \textbf{Data availability:} No data associated with the manuscript.
\item \textbf{Author contributions:}  A.K.N. contributed to the code development and performed the simulations. A.K.N. and A.S. involved in the data analysis. M.M. contributed to research design, and M.S.S. proposed the numerical development strategy, designed the research topic and plans, as well as involved to the funding acquisition, resource provision, and project administration. All authors contributed to the writing and interpretation of the article.
\end{itemize}

\end{document}